# Many-Body Basis Set Amelioration Method for Incremental Full Configuration Interaction

Jeffrey Hatch, Alan E. Rask, Duy-Khoi Dang, Paul M. Zimmerman*
Department of Chemistry, University of Michigan
930 N. University Ave, Ann Arbor, MI 48109, USA, *paulzim@umich.edu

**Abstract**:
Incremental full configuration interaction (iFCI) is a polynomial-cost electronic structure method that systematically approaches the FCI limit by employing the method of increments to solve the Schrödinger equation through a many-body expansion. This article introduces the many-body basis set amelioration (MBBSA) method, which is designed to allow iFCI to be applicable to larger atomic orbital basis sets. MBBSA uses a series of inexpensive iFCI calculations to approximate the correlation energy that would be found using a more expensive, highly accurate iFCI calculation. When compared to standard iFCI computations on smaller molecules in triple-zeta and larger basis sets, MBBSA provides approximations to the total and relative energies within chemical accuracy. MBBSA exhibits a reduced cost of between 60-92% when compared to standard iFCI calculations, with larger systems experiencing the largest benefit. Tests of MBBSA on two reactions that involve highly correlated systems, the automerization of cyclobutadiene and a Criegee intermediate reaction, show that MBBSA has practical utility for studying realistic chemistries.

**Introduction:**
Electronic structure methods are computational approaches that give great insight into a variety of complicated molecular species by approximately solving the electronic Schrödinger equation.[1–18] Wave function theory (WFT) can achieve qualitative and quantitative accuracy for weakly- and strongly-correlated systems alike. WFT also has the appealing feature of being systematically improvable, albeit with quickly increasing costs as the level of sophistication grows. As a result, wavefunction methods can be roughly arranged hierarchically along a ladder of simultaneously increasing cost and accuracy.[19–21] Due to the tradeoffs between cost and accuracy, it becomes important for each WFT methodology to employ strategies to mitigate costs without sacrificing accuracy.[22,23]

Electronic structure methods are typically expressed in a carefully chosen, finite basis set. Atom-centered, single-particle basis functions have played a critical role in approaching accurate WFT solutions in a systematic way, while maintaining good tractability as the basis size increases.[24–28] Correlation consistent atomic orbital basis sets have been used to methodically increase the flexibility of the basis. For instance, the cc-pVXZ family of basis sets (correlation consistent, polarized valence X-zeta; X refers to double-zeta, triple-zeta, etc), gives a recipe for the number of basis functions and polarization functions at each basis rank. As the basis size grows (higher zeta) the total energy of a WFT approaches its basis limit, with less and less additional correlation being recovered with each step up in the basis size. This asymptotic convergence allows results from a series of correlation consistent basis sets to usefully extrapolate to a complete basis set limit.[29–39] Many extrapolation methods have been proposed over the years with varying efficacy and accuracy, utilizing differing numbers of points in the extrapolation and different functions to anticipate convergence.[40–46]

The electronic Schrödinger equation can be solved exactly in a given basis set by the method known as full configuration interaction (FCI).[47–49] FCI wave functions are exact because they contain every possible electron configuration available within the basis, though this combinatorial number makes FCI impractical for all but the smallest systems.[50,51] However, FCI has been shown to produce mostly "deadwood" configurations, meaning that most configurations contribute an insignificant amount to the total wave function.[52,53] As a result, a number of methods have been developed in an attempt to eliminate as much deadwood as possible, while still maintaining the accuracy of FCI. In recent years, the most popular methods of this type are select CI approaches, which algorithmically determine the most important configurations in FCI.[54–61] These techniques lower the computational cost of FCI by orders of magnitude but are still plagued by exponential scaling. Another approach uses the many-body expansion of increasingly complex n-body terms.[62,63] This approach has been explored independently in many-body expanded FCI and incremental FCI (iFCI), the latter of which is the focus of this paper.[64,65] iFCI has the same goal as other select CI methods, but a qualitatively different strategy to approach the exact FCI solution. iFCI expands the correlation energy via an incremental expansion of increasingly complex $n$-body terms. As more terms are included, iFCI approaches the exact energy, allowing systematic addition of the most important configurations while limiting the computational cost.[66–68] This expansion results in polynomial cost, giving it a key advantage over related approaches. This low order scaling has allowed iFCI to capture dynamic and static correlation in a variety of systems of varying size and complexity, including strongly correlated small molecules, diradicals, transition metal complexes and per- and polyfluorinated chains (PFAS).[66,67,69–71]

Even with the key advantage of polynomial scaling, iFCI remains a relatively costly method due to the need to solve the numerous terms emerging from the *n*-body expansion. This is especially the case as the size of the basis grows larger. iFCI therefore utilizes a few techniques to mitigate cost: an improved reference state from the perfect pairing wavefunction[72–75], utilization of the natural orbitals to truncate the virtual space, and a fast CI solver, heat-bath CI (HBCI).[16,61,76–80] Even with these improvements, iFCI has been limited to triple-zeta basis sets for main-group molecules and double-zeta basis sets for transition metal complexes.[66,67,70] In most cases polarized double-zeta basis sets are insufficient to reach chemical accuracy, and general wisdom suggests that basis sets need to be at least polarized triple-zeta in quality.[76,81–83] To mitigate costs and achieve larger-basis-like results, some electronic structure methods have combined the results of high accuracy methods in a small basis with lower accuracy methods in a large basis. These strategies include the Gaussian-n methods (G-n) and the correlation consistent composite approach (ccCA).[84–88] These all rely on one key insight: that the results of a high accuracy electronic structure calculation can be imitated via a series of less accurate but more tractable calculations.

Herein we will take inspiration from G-n and ccCA approaches to enable iFCI to work with larger basis sets. Consider two levels of wavefunction theory, a low-level theory (LLT) and a high-level theory (HLT). HLT is more accurate (and costly) than LLT. Also consider two basis sets, a larger basis (LB) and a smaller basis (SB), with the former being more accurate but costly. Therefore, computing HLT in SB is plausible (hitherto: HLT(SB)), but not HLT in LB (HLT(LB)). Both basis sets are amenable to LLT.

$$\Delta E_{LLT} = E_{LLT(LB)} - E_{LLT(SB)} \tag{1}$$

In an additive correction scheme, Eqn. 1 represents the missing correlation energy due to basis set incompleteness (using LLT). Presuming $\Delta E_{HLT}$ has little-to-no dependence on the level of theory (a realistic but not perfect assumption), an accurate approximation to $E_{HLT(LB)}$ would be Eqn. 2.

$$E_{HLT(LB)} = E_{HLT(SB)} + \Delta E_{LLT} \qquad (2)$$

No computation of $E_{HLT(LB)}$ is required. A possible composite scheme of this type would be constructed from two basis sets (e.g., SB = DZ and LB = TZ) and two levels of theory. iFCI, having a well-defined hierarchy of correlation, can be used as a composite approach where the levels LLT and HLT would be $n-1$ and $n$, where $n$ is the ultimate target level of correlation. A composite approach with iFCI is thus defined by the combination of basis sets and their corresponding order of the n-body expansion.

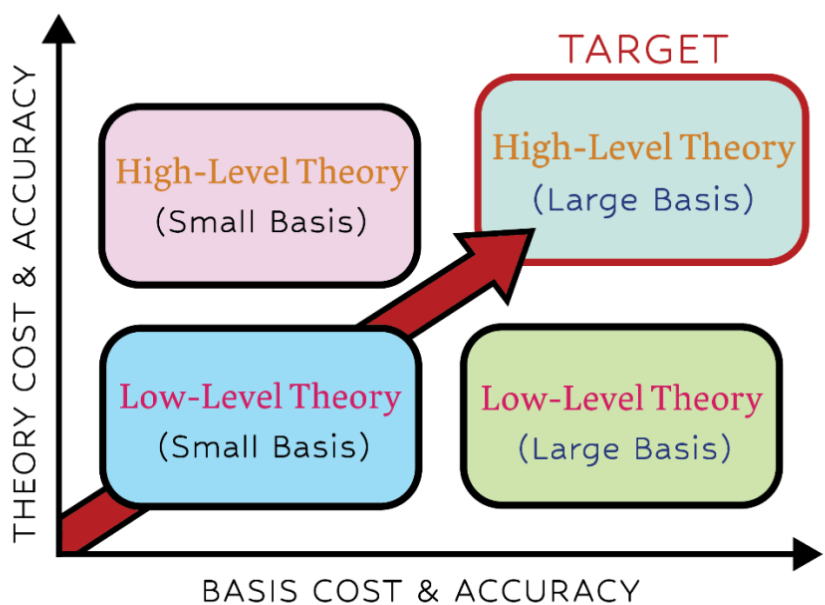

Figure 1. Schematic showing how the results of a high-level theory (HLT) in a large basis (LB) can be approximated as a combination of calculations in lower levels of theory (LLT) and/or smaller basis sets (SB).

This work introduces the many-body basis set amelioration (MBBSA) method to increase the range of applicability of iFCI, allowing estimates of a larger basis set results at greatly reduced cost (see figure 1). The MBBSA method specifically avoids computing the correlation energy of $n$-body iFCI with a large basis, substituting instead large basis $(n-1)$-body iFCI and small basis $n$-body iFCI computations. To show the range of utility of this approach, this method is first applied to calculate the total energies of a few polyatomic systems. Subsequently, tests on a set of isodesmic reactions will show MBBSA's effectiveness in calculating thermodynamic properties.[89,90] A hydrogen transfer barrier height (HTBH) reaction set provides a more challenging test involving activation energies and thermodynamic properties of open-shell and highly correlated species.[91,92] Finally, the MBBSA method is applied to the automerization of cyclobutadiene and a Criegee intermediate reacting with ethylene. These two reactions use the MBBSA method to find activation energies and thermodynamic energies for strongly correlated chemical species, a true test for MBBSA.

## Methods:

### The Many-Body Expansion of iFCI:

The many-body expansion of FCI provides a size extensive, polynomial-cost approximation to FCI that also forms the starting point for the MBBSA method. In iFCI, the correlation energy is retrieved via a series of CI computations involving increasingly larger groups of electrons. Specifically, at the $n$-body level, $2n$ electrons are correlated in each grouping, and the sum across all groupings yields the total correlation energy. In this article the iFCI energy is expressed as

$$E = E_{ref} + \sum_i \epsilon_i + \sum_{i<j} \epsilon_{ij} + \sum_{i<j<k} \epsilon_{ijk} + \cdots \quad (3)$$

where

$$\epsilon_i = E_c(i)\big|_{\zeta_i} \quad (4)$$
$$\epsilon_{ij} = E_c(ij) - E_c(i) - E_c(j)\big|_{\zeta_{ij}} \quad (5)$$
$$\epsilon_{ijk} = E_c(ijk) - E_c(ij) - E_c(ik) - E_c(jk) - E_c(i) - E_c(j) - E_c(k)\big|_{\zeta_{ijk}} \quad (6)$$

and the indices, i, j, k… refer to the n-bodies of the expansion, each representing the two electrons of an occupied orbital. For example, at the 2-body level, 4 electrons at a time are correlated via CI computations within an active space of $2n + N_v$ orbitals. The $\zeta_{ij}$ term refers to the $N_v$ natural orbitals (NOs) which constitute the virtual space that the $2n$ electrons are correlated within. These NOs are formed by diagonalizing an approximate density matrix for each $ij$ 2-body term, as described in the computational details section. To avoid overcounting correlation from duplicate configurations generated at higher-n, the correlation energies of the $i$ and $j$ 1-body terms are computed using the same NOs and subtracted from the 2-body term (c.f. Eqn. 5). Similar considerations apply for any term with $n > 1$ (Eqn. 5, 6 and similar expressions for higher n). This procedure, referred to as the incremental natural orbital (iNO) approach, is designed to facilitate systematic convergence of the $n$-body expansion.[93]

The bodies for iFCI in this work are defined using the perfect pairing (PP) procedure, which gives localized bonding orbitals and their antibonding counterparts.[72–75] PP orbitals have been shown to be useful in capturing some static correlation in the reference state (c.f. Eqn. 3), and also allowing

$\epsilon_X$ values to decrease with order of expansion, in particular past the 2-body level (e.g., $|\epsilon_{ijk}| < |\epsilon_{ij}|$). Further details of the iFCI approach are available in refs. 66-70.

Results from prior studies suggest that iFCI nearly eliminates the method error, as it approaches the FCI limit in a given basis. For example, the correlation energy of benzene computed via iFCI at the 4-body level agreed with other high-level methods to within about 1 mHa.[70] Under the assumption the iFCI is converged at the 3- or 4-body levels, the finite size of the atomic orbital basis set is the main limitation of iFCI.

**Composite Methods:**

The MBBSA method draws inspiration from related strategies known as composite methods. Two well-known examples are the Gaussian-n (n being 2, 3, 4) family of methods and the correlation consistent Composite Approach (ccCA).[94–96] Each of these methods uses a series of lower-level-of-theory computations with increasing basis set size to describe how the electronic energy depends on basis. This basis extrapolation is then combined with higher-level-of-theory computations in relatively smaller basis sets to describe the effect that level of theory has on the correlation energy. For example, the Gaussian-n family attempts to reproduce coupled cluster single, double, with perturbative triples (CCSD(T)) results in a large basis, without performing the large basis CCSD(T) computation. These procedures seek to systematically reduce the basis error and the method error while not incurring the full cost of the high-level, high-basis computation.

Related to composite methods are dual basis techniques that use two basis sets to reduce cost and maintain accuracy. The idea was first introduced by Jurgens-Lutovsky and Almlöf[97] who applied this concept to MP2 by first using different basis sets for Hartree-Fock and higher stages of correlation to reproduce traditional MP2 results. The dual basis method was then combined with incremental techniques by Dolg et al [85,98] to approximate results from CCSD and CCSD(T). Utilizing the inherent parallelization of their incremental method, they were able to treat much larger systems than would be possible via canonical CCSD(T).

**The MBBSA Method**

The many-body basis set amelioration (MBBSA) method approaches the large basis FCI limit through a sequence of iFCI computations. MBBSA requires a pair of computations: iFCI through order $n$ with a smaller basis and iFCI through $n-1$ with a larger basis. MBBSA helps to address finite basis set errors in iFCI, while retaining the favorable properties of iFCI, such as size extensivity and low method error. Since iFCI with $n \geq 3$ captures static and dynamic correlation but becomes intractable with larger basis sets, our focus here is on achieving larger-basis, iFCI-level results with $n = 3,4$.

To motivate MBBSA, an estimate of the cost of iFCI computations can be made by considering the number of electron configurations (i.e., determinants) in each incremental term. For an increment involving $n_\alpha + n_\beta$ electrons and $N$ active orbitals, the number of determinants is $\binom{N}{n_\alpha}\binom{N}{n_\beta}$. When the system has equal numbers of $\alpha$ and $\beta$ electrons, this simplifies to $\binom{N}{n}^2$, where $n$ is the number of electron pairs in occupied orbitals. The number of determinants in an increment then scales as $N^{2n}$, meaning that the number of determinants at the 3-body level would be $O(N^6)$, or $O(N^8)$ at the 4-body level. The iFCI algorithm therefore inherently needs a method to reduce the number of active orbitals present in any increment to avoid becoming intractable. A natural-orbital-based screening procedure is used[68] to limit the number of virtual orbitals in each increment. This strategy selects virtual orbitals that tend to be localized near the occupied active orbitals, so in practice, $N$ does not grow quickly with system size. Rather, $N$ grows quickly with basis set size.

Previous work has shown that the iFCI expansion adds less correlation energy per increment as the n-body level increases past $n = 2$.[66–70] Herein, we demonstrate that the correlation energy at the $n = 3$ or $n = 4$ levels does not vary strongly with the size of the basis. MBSSA exploits this consistency to approximate the highly expensive calculation of the *n*-body correlation energy. The n-body correlation energy is approximated by combining the much cheaper (n-1)-body correlation energy with the correlation energies of the n and (n-1)-bodies in a smaller, less costly basis set. This will be shown to converge on the FCI limit at significantly reduced cost with minimal loss in energetic accuracy.

The equation associated with the MBBSA method as it relates to iFCI in this work is

$$E^{iFCI}[n]_{LB} \approx E^{iFCI}[n-1]_{LB} + \left(E^{iFCI}[n]_{SB} - E^{iFCI}[n-1]_{SB}\right) \quad (7)$$

where SB and LB refer to small and large basis sets, respectively. In this study, n refers to the 3-body level in most instances and the 4-body level otherwise. It has been suggested that a triple zeta (TZ) basis is the minimum size required to reach chemical accuracy for small to medium-sized systems.[99] As such we considered a polarized TZ basis to be appropriate as the LB for this study, though larger basis sets are plausible as well. For small molecules ($H_2O$, $CH_2$, and $CH_4$) we tested MBBSA in 3 different combinations of basis sets. In this case, LB was the cc-pVXZ basis and SB was the cc-pV(X-1)Z basis where X corresponds to the basis set hierarchy: D (double-zeta), T (triple-zeta), Q (quadruple-zeta), and 5 (quintuple-zeta). From this point on the MBBSA correction where $n = 3$ in Eq. (7) will be referred to as MBBSA-3 and where $n = 4$ in Eq. (7) will be called MBBSA-4.

**<u>Computational Details:</u>**

All computations were performed in a development version of the QChem software package.[100] Perfect pairing (PP) orbitals were formed from Pipek-Mezey or Boys (See supplemental Information) localization of the Hartree-Fock ground states to calculate the reference energy and the molecular orbitals used in the iFCI procedure, followed by Sano determination of initial virtual orbitals and full orbital optimization under the pairing ansatz.[101–108] Geometries for each of the molecules from the HTBH reaction set were taken from the database itself, while all others were optimized using the resolution-of-the-identity MP2 approach (RIMP2) and the cc-pVTZ basis[109] combined with the RIMP2-cc-pVTZ auxiliary basis.[110]

iFCI computations were performed up to the $n = 4$ level, as specified in the results that follow. For each *n*-body term, a heat bath configuration interaction (HBCI) solver was used to produce the correlation energy $E_c(X)$. This method is discussed extensively in refs 16, 61 and 76-80. HBCI depends on convergence parameters called $\varepsilon$, which control HBCI's approach to the FCI limit. Herein, $\varepsilon_1$ was set to 0.5 mHa, and $\varepsilon_2$ to 0.1 µHa, which correspond to the variational and perturbative steps, respectively. iFCI also utilizes a convergence parameter ($\zeta$) which controls inclusion of virtual NOs in each incremental term (see ref 68). This is the same value as in Eqn. 3-6. The $\zeta$ values are reported in the SI.

NOs are generated by diagonalizing the 1-particle reduced density matrix. The SCF wavefunction does not change with respect to orbital rotation but finding the optimal orbital orientation to facilitate the treatment of electron correlation has been of interest for many years.[111] NOs are generally considered to be a good choice for representing the one-particle basis functions as they generally require the fewest determinants for fixed accuracy compared to other methods for orbital selection.[112] The convergence of iFCI on the FCI limit has been shown to depend on the initial orbital choice with PP orbirtals generally outperforming HF orbitals. It will be shown that convergence is also aided by the use of NOs in the form of the iNO approach.

While testing the MBBSA method, it was observed that the occupied molecular orbitals in the SB must be qualitatively the same as the LB to ensure smooth convergence. The PP procedure, however, often produces multiple local minima dependent on qualitatively different orbital localization. Therefore, PP orbitals were rotated and then reoptimized as needed to ensure alignment of the orbitals between PP/SB and PP/LB computations.

## Results and Discussion:

This section will showcase MBBSA's versatility by applying it to representative test cases. First, the absolute energies of a few small molecules will be analyzed, illustrating the cost saving power of MBBSA compared to standard iFCI. Then, a set of isodesmic reactions will test how accurately MBBSA can estimate thermodynamic properties. Next, thermokinetic properties of reactions in the HTBH database, which contains highly correlated small to medium open- and closed-shell molecules, will be studied through the lens of MMBSA. Finally, MBBSA will be applied to a reaction involving a Criegee intermediate and to the automerization of cyclobutadiene. These represent difficult test cases for our method as they involve highly correlated molecular species of moderate size.

### Small Molecule Energy Convergence and Computational Cost:

Methylene ($CH_2$), methane ($CH_4$), and water ($H_2O$) were considered, with six, eight, and eight electrons being correlated, respectively. The MBBSA-3 method was used for $CH_2$ and the MBBSA-3 and MBBSA-4 methods for $CH_4$ and $H_2O$. MBBSA-3 is approximating the 3-body energy in the larger basis and MBBSA-4 is approximating the 4-body energy in the larger basis. Each molecule was examined using a series of correlation-consistent basis sets from the polarized, double-zeta level through the polarized, pentuple-zeta level: cc-pVDZ, cc-pVTZ, cc-pVQZ and cc-pV5Z. Errors were computed as the difference between the MBBSA energy and the standard iFCI energy. For each benchmark, the MBBSA value is compared to the iFCI(n=3) calculation in the large basis.

**Table 1.** Errors associated with approximating the high-level theory (HLT) in the larger basis (LB). In each case the low-level theory used in the MBBSA method was the (n-1)-body and the small basis was (X-1)Z.

|          |       | Error in Approximating HLT(LB) (mHa) | | |
| -------- | ----- | -------- | -------- | -------- |
| Molecule | HLT   | LB = TZ  | LB = QZ  | LB = 5Z  |
| Methylene | $n = 3$ | 0.147 | 0.075 | 0.041 |
| Water    | $n = 3$ | 1.068 | 0.733 | 0.175 |
| Water    | $n = 4$ | 0.048 | 0.127 | 0.064 |
| Methane  | $n = 3$ | 0.908 | 0.416 | 0.166 |
| Methane  | $n = 4$ | 0.225 | 0.038 | 0.010 |

Relative energies in Table 1 indicate that errors with MBBSA are very low, being below 1.1 mHa (0.7 kcal/mol) in each case. The errors decrease with increasing $n$-body level and mostly decrease with increasing basis size. One case shows a deviation from this trend, where the largest error in water at $n = 4$ is when the large basis (LB) is the QZ basis. However, the $n = 4$ errors for TZ, QZ, and 5Z are each below 0.16 mHa,, indicating satisfactory convergence of the MBBSA procedure.

Having shown MBBSA can reproduce total energies to good accuracy, the computational costs of MBBSA were examined. As a representative example, Figure 2 illustrates the computational cost of iFCI with respect to the number of basis functions and $n$-body level for the water molecule. This figure makes clear that the most expensive part of FCI is the highest $n$-body level, with the lower $n$ levels requiring substantially less compute time. Because the computational cost of the highest $n$ level also scales poorly with basis set size (see Methods Section), this cost is compounded in larger basis sets.

The MBBSA approach approximates the iFCI (n=4)/LB result without computing the 4-body/LB set of increments. Therefore, it is useful to compare the total cost of MBBSA to the cost of the 4-body level in a large basis. As shown in Figure 2b, the 4-body LB incremental terms require more than twice the computational time as the entire MBBSA procedure, indicating a significant cost reduction even for systems with as few as eight valence electrons.

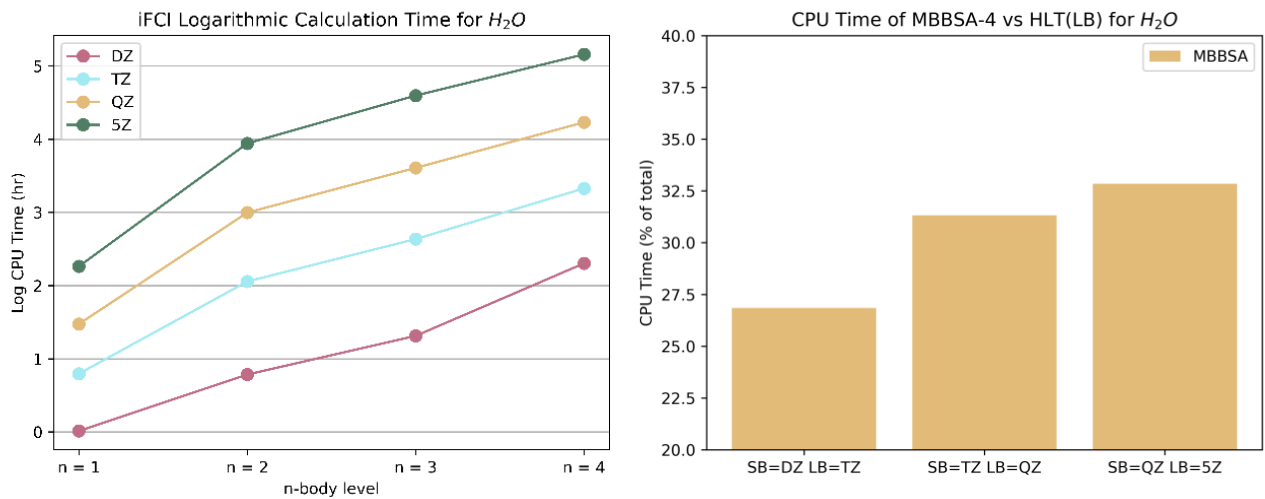

(a) (b)

Figure 2. The CPU time as a function of *n*-body level and basis set size for water. a) The cost associated with the cc-pVXZ basis with the RIMP2-cc-pVXZ basis sets where X represents D, T, Q, and 5. b) The cost associated with the MBBSA method in a series of large and small basis sets. In each case the sum of the cost of the $(n = 1 - 4)$-body levels in the small basis and the $(n = 1 - 3)$-body levels in the large basis is exrpressed as a percentage of the cost of the 4-body level in the large basis.

**Isodesmic Reaction Set:**

Isodesmic reactions, which preserve the number and type of each bond between reactants and products, were considered as the next test set for MBBSA.[113] These reactions involve closed-shell, small to medium-sized molecules ranging in size from methane with 10 electrons to a pyran derivative that has 46 electrons.[114] The 11 isodesmic reactions in Table 2 include reactants and products with rings and/or multiple lone pairs on a single atom, making them nontrivial cases for any correlated method.[115,116] As such, this test set can delineate the theory error from the basis set errors through the lens of the MBBSA approach.

**Table 2**. The unsigned error of the thermodynamics of the 11 isodesmic reactions in the MBBSA-3 method using the cc-pVTZ and cc-pVDZ basis sets. Theory error represents the difference between HLT and LLT in each basis set with the difference between the theory errors representing the unsigned MBBSA error as reported in the last column. All errors in kcal/mol.

| Reaction | Reactants | | Products | | Theory Error (DZ) | Theory Error (TZ) | Unsigned Error |
|---|---|---|---|---|---|---|---|
| 1 | $NH_2CH_2CHCH_2$ | $C_2H_6$ | $NH_2CH_2CH_2CH_3$ | $C_2H_4$ | 1.46 | 1.12 | 0.34 |
| 2 | $CH_2CHCH_2Cl$ | $CH_4$ | $ClCH_2CH_3$ | $C_2H_4$ | 1.57 | 1.43 | 0.13 |
| 3 | $CH_3CH(NH_2)CH_3$ | $CH_4$ | $NH_2CH_2CH_3$ | $C_2H_4$ | 2.49 | 2.35 | 0.13 |
| 4 | $C_2H_3N$ | $C_2H_6$ | $C_2H_5N$ | $C_2H_4$ | 0.91 | 0.71 | 0.20 |

| 5 | $C_2N_2H_2$ | $C_2H_6$ | $C_2N_2H_4$ | $C_2H_4$ | 0.05 | -0.39 | 0.43 |
|---|---|---|---|---|---|---|---|
| 6 | $C_3NH_5$ | $C_2H_6$ | $C_3NH_7$ | $C_2H_4$ | 6.99 | 6.43 | 0.56 |
| 7 | $C_4NH_5$ | $C_2H_6$ | $C_4NH_7$ | $C_2H_4$ | 14.35 | 13.27 | 1.08 |
| 8 | $C_5OH_6$ | $C_2H_6$ | $C_5OH_8$ | $C_2H_4$ | 2.58 | 2.69 | 0.11 |
| 9 | $FCH_2CH_2CH_3$ | | $CH_3CH_2(F)CH_3$ | | -4.63 | -4.37 | 0.30 |
| 10 | $C_4H_5Cl$ | $C_2H_6$ | $ClCH_2CH_3$ | $C_4H_6$ | 2.34 | 1.96 | 0.37 |
| 11 | $NH_2CH_2Cl$ | $C_2H_6$ | $CH_3Cl$ | $NH_2CH_2CH_3$ | 1.33 | 1.21 | 0.12 |
| MUE | | | | | 3.52 | 3.26 | 0.34 |

The results of the isodesmic reaction analysis from MBBSA-3 are provided in Table 2. The theory error (equation 6) in Table 2 is the difference between the iFCI reaction energy at the 3- and 2-body levels, respectively:

$$Theory\ Error = \Delta E_{iFCI}[n = 3] - \Delta E_{iFCI}[n = 2] \qquad (8)$$

The unsigned error of Table 2 represents the deviation of the MBBSA-3 reaction energy from the $E^{iFCI}[n = 3]_{TZ}$ results. The last row of the table shows mean unsigned error (MUE) across the entire reaction set.

Generally, the theory errors show that the 2-body level of iFCI does not reach chemical accuracy, with errors greater than 1 kcal/mol. Table 2 shows that reactions 6 and 7, which involve 4-membered rings, have by far the largest theory errors. iFCI can accurately describe these transformations, but only when $n > 2$, as has been the case in prior studies involving iFCI.[66–70] The 3-body level of correlation is required to reach chemical accuracy, but notably, the theory errors with the two basis sets are similar, and thus cancel in the MBBSA energies. At the MBBSA-3 level of theory, the MUE is 0.34 kcal/mol, with a maximum error of 1.08 kcal/mol. MBBSA-3 therefore forms an effective approximation of polarized triple-zeta, FCI-quality results for the isodesmic reaction set.

**Hydrogen Transfer Barrier Height (HTBH) Reaction Set:**
The hydrogen transfer barrier height (HTBH) reaction set consists of activation energies and thermodynamic properties of small to medium-sized molecules and presents a challenging test for any correlated approach.[91,92] This reaction set introduces two things that have not been considered in the prior examples of this work, namely, transition states and open-shell molecules. The open-shell transition states are particularly challenging due to the stretched bonds present, having delocalized electronic structures where static as well as dynamic correlation must be accounted for to attain quantitative accuracy. In each case, the MBBSA relative energies are compared to the *n*-body energies in the large basis. The errors in Figure 3 represent the MBBSA-3 and MBBSA-4 deviations from the standard iFCI *n*-body energy. Two of the 18 reactions in the data set were chosen as representative examples of the MBBSA with the remaining errors reported in the SI. Reaction 8 (yellow in Figure 3) was the most challenging reaction for MBBSA with the correspondingly highest errors while reaction 15 (Blue in Figure 3) represents a more average example that aligns well with the MUE for the test set.

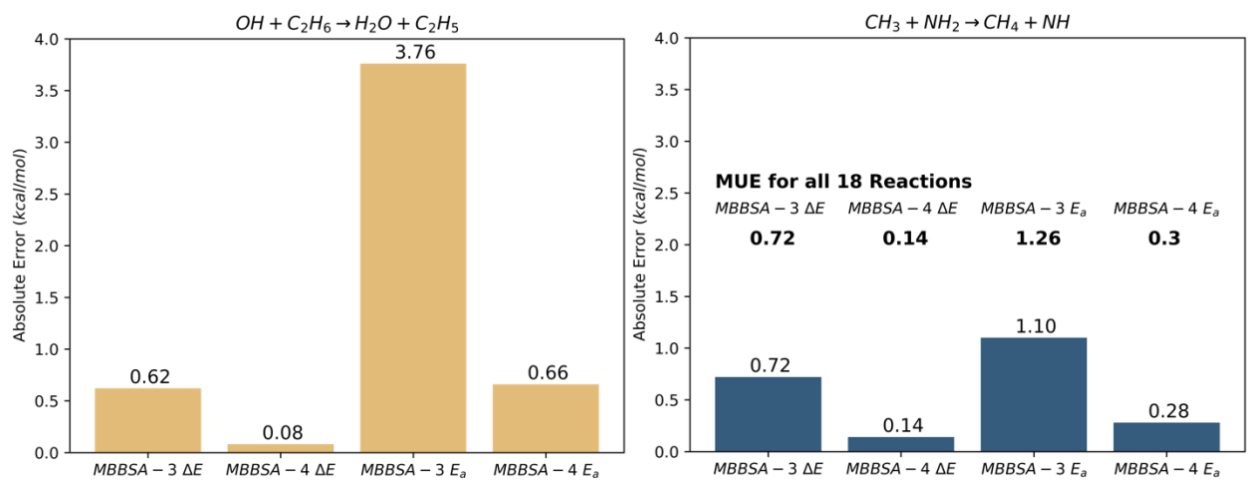

Figure 3. Two representative reactions showing the error of MBBSA-3 and MBBSA-4 in predicting the thermodynamics (ΔE) and the activation energy ($E_a$) quantities along with MUEs for all reactions in the data set. The reaction with the largest error (yellow) can still achieve the threshold for chemical accuracy using MBBSA-4.

Finding the $E_a$ represents a more challenging problem than ΔE due to the inclusion of transition state energies. As a result, the $E_a$ error in Figure 3 is larger than the ΔE error for a specific reaction at a given level of theory. The MUEs associated with MBBSA-3 in this reaction set are 1.26 kcal/mol for $E_a$ and 0.72 kcal/mol for ΔE. In contrast, individual MBBSA-4 errors are all within chemical accuracy, indicating that increasing the $n$-body level improves convergence. The MUE in $E_a$ for MBBSA-4 was 0.30 kcal/mol, indicating good performance across these challenging open-shell reactions.

**Automerization of Cyclobutadiene and Criegee Intermediate Reaction:**
In this section, additional reactions involving small molecules are studied. Both examples involve polyatomics with significant dynamic correlation, and varying amounts of static correlation along their respective reaction coordinates. These provide challenging test cases with relevance to real-world chemical applications where FCI is needed.

Cyclobutadiene automerizes from a $D_{2h}$ geometry through a $D_{4h}$ transition state (see Figure 4).[117] The frontier π orbitals are nearly degenerate at the $D_{2h}$ structure and exactly degenerate at the $D_{4h}$, making for biradical character that is challenging to simulate. Activation barriers ($E_a$) computed with various electronic structure theories suggest the need for at least a triple-zeta, polarized basis set (Figure 4). The highly correlated nature of this system also yields significant discrepancies in the activation energy. SF-DFT calculations generated a range of activation energies, ranging over nearly 9 kcal/mol across a set of 8 density functionals. Wave function methods CCSD and CCSDT differ by approximately 1 kcal/mol, which is the same size as the difference in barriers from

CASPT2 with two different active spaces. To refine the barrier further, closer approaches to the FCI limit are warranted.

Figure 4 shows that activation energies from MBBSA-3 and iFCI(n=3) agree well with one another and with the extrapolated FCI energy from Ref. 117. The Ea values also fall within the experimental range of barriers, though this quantity is highly uncertain. [118–121] The MBBSA-3 error compared to iFCI(n=3) is 0.43 kcal/mol, correcting for a little more than half of the basis set error at the cc-pVDZ level.

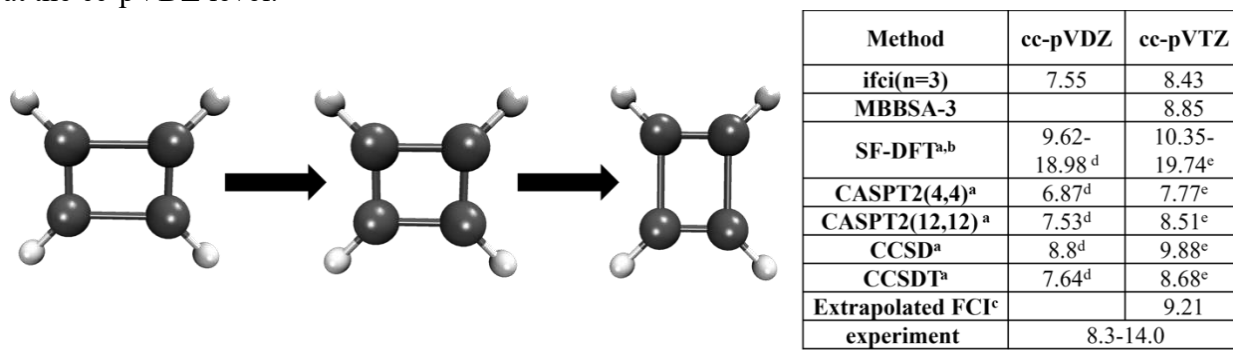

| Method | cc-pVDZ | cc-pVTZ |
|---|---|---|
| ifci(n=3) | 7.55 | 8.43 |
| MBBSA-3 | | 8.85 |
| SF-DFT[a,b] | 9.62-18.98[d] | 10.35-19.74[e] |
| CASPT2(4,4)[a] | 6.87[d] | 7.77[e] |
| CASPT2(12,12)[a] | 7.53[d] | 8.51[e] |
| CCSD[a] | 8.8[d] | 9.88[e] |
| CCSDT[a] | 7.64[d] | 8.68[e] |
| Extrapolated FCI[c] | | 9.21 |
| experiment | 8.3-14.0 | |

Figure 4. The automerization of cyclobutadiene ($C_4H_4$) from a $D_{2h}$ to $D_{4h}$ to $D_{2h}$ geometries. Inset shows activation energies in kcal/mol. a) Ref. 119. b) SF-DFT range from 8 functionals, see Ref. 119. c) Ref. 118 d) aug-cc-pVDZ basis. e) aug-cc-pVTZ basis.

The effectiveness of the MBBSA-3 method can be traced to the behaviors of individual correlation energies in the 3-body expansion for cyclobutadiene. Figure 5 shows the magnitude of the 3-body terms and the cumulative correlation energy for each basis and geometry. The profiles follow each other closely across the two basis sets. In addition, because some 3-body terms have errors in the positive direction and others in the negative, cancellation in errors appears as the sum is completed. The small deviations and error cancellation eventually result in the 0.4 kcal/mol MBBSA-3 error compared to the full iFCI(n=3) energy. The SI (Figure S2) shows that such convergence is not possible at the 2-body level, due to the significant differences in correlation energies (large theory error).

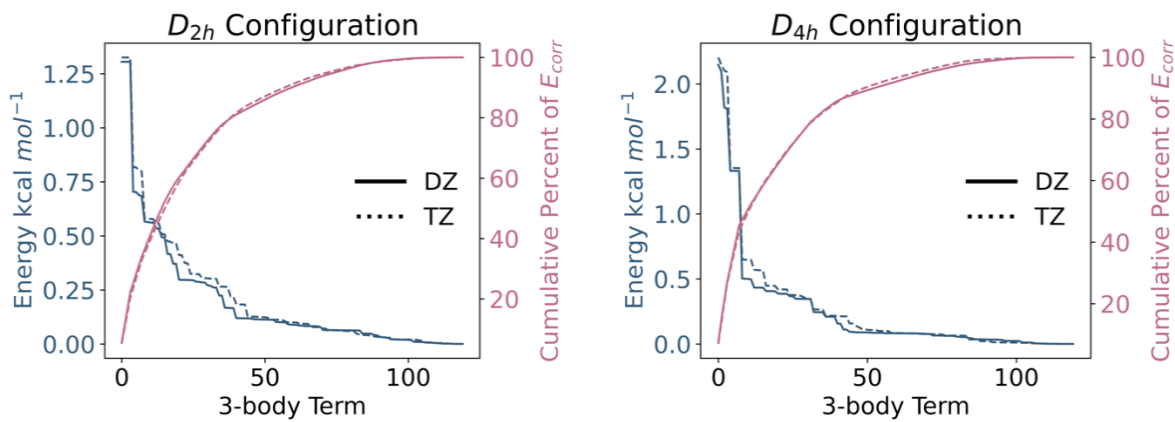

Figure 5. A comparison of the magnitude of correlation energy in each basis and the cumulative percent of the total correlation energy recovered by each 3-body term.

Criegee intermediates are zwitterionic biradical species that are important nighttime oxidants in the atmosphere and have been implicated in secondary organic aerosol formation.[122] The mixed electronic characteristics (strong correlation) of these molecules makes them a difficult test case for the MBBSA method. One example is the reaction involving a Criegee derivative shown in Figure 6.[123] To give an idea of the uncertainty in applying conventional electronic structure methods to such Criegee reactions, the inset within Figure 6 shows thermokinetic results from density functional theory. With a triple-zeta basis, the BLYP and B3LYP functionals provide activation energies separated by 6 kcal/mol. Therefore, MBBSA was applied to provide a more systematic approach to revealing the reaction's characteristics.

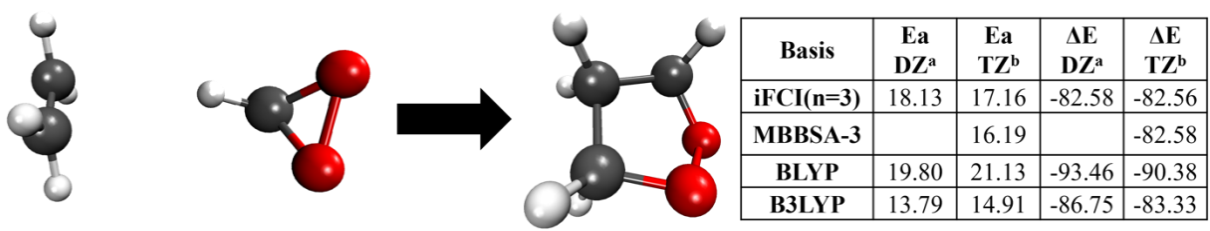

| Basis | $E_a$ DZ[a] | $E_a$ TZ[b] | $\Delta E$ DZ[a] | $\Delta E$ TZ[b] |
|---|---|---|---|---|
| iFCI(n=3) | 18.13 | 17.16 | -82.58 | -82.56 |
| MBBSA-3 |  | 16.19 |  | -82.58 |
| BLYP | 19.80 | 21.13 | -93.46 | -90.38 |
| B3LYP | 13.79 | 14.91 | -86.75 | -83.33 |

Figure 6. The reaction of dioxirane (a Criegee derivative) and ethylene to form a 5-membered ring (1,2-dioxalane). Inset shows activation energy and reaction energy in kcal/mol. a) cc-pVDZ basis. b) cc-pVTZ basis.

In this highly exothermomic reaction, the O-O bond of the dioxirane molecule breaks to form a biradical intermediate which reacts with $C_2H_4$ to form the 5-membered ring. The activation energy (approximately 17 kcal/mol) is moderate but surmountable under atmospheric conditions. The MBBSA-3 error in the activation energy is 1.08 kcal/mol and the error for the reaction energy was 0.02 kcal/mol. MBBSA-3 therefore provides a good approximation to iFCI close to chemical accuracy.

With the Criegee reaction, electronic structure simulations must treat 11 atoms and 30 valence electrons. In the cc-pVTZ basis set, these correspond to a FCI dimension of $10^{46}$. Compared to the essentially intractable FCI problem, iFCI and MBBSA require computations of 15 one-body terms, 105 2-body terms and 455 3-body terms, with maximum CI dimension of $10^{13}$. The iFCI(n=3) calculations are already a dramatic simplification compared to FCI, but still took more than 95 hours of CPU time. The MBBSA calculations took only 6.7 hours of CPU time. MBBSA therefore results in a cost reduction of 92% compared to the iFCI computation and brings the total compute time down to reasonable levels for a correlated electronic structure calculation. Additionally, we anticipate that the further cost reduction that scales with an increase in system size will prove to be invaluable when applying MBBSA to larger systems in the future. Since MBBSA brings even larger fractional cost savings as the system size grows, we anticipate its value to hold for larger chemical species that will be examined in the near future.

## Conclusion

iFCI has previously been shown to simplify the FCI problem and reduce its costs to tractable levels, all the while recovering the full spectrum of static and dynamic correlation. MBBSA takes this route further, allowing larger basis sets to be used along iFCI without the concomitant increase in cost. Compared to iFCI, examples show that MBBSA cost savings range from ~60% for small cases like water to 92% for medium sized reactions involving the Criegee intermediate. The ability to confidently approximate these high accuracy methods at a fraction of the cost increases our ability to target larger and more complex chemical systems.

Beyond the cases studied here, MBBSA has the potential to extend iFCI's reach into areas previously considered infeasible. Future applications could include modeling transition metal reactivity in catalysis, where the combination of strong correlation and large basis sets presents a persistent challenge. Similarly, high-accuracy treatment of large conjugated π-systems—relevant to organic electronics and photochemistry—may now be within reach. The scalability of MBBSA also suggests a pathway toward systematic benchmarking of machine-learned density functionals, providing near-exact data for training sets at a fraction of the usual cost. With continued refinements and optimizations, MBBSA could become a key tool in pushing the limits of electronic structure theory toward larger and more complex chemical environments.

## Acknowledgments
The authors would like to thank the U.S. Department of Energy, Office of Science, Basic Energy Sciences for their support of this work through DE-SC0022241. Our thanks also to David Braun who continuously supports our computational endeavors.

## Supporting Information
The Supporting Information is available free of charge at link-to-be-determined

The supporting information contains:
- The iFCI settings ($\zeta, \epsilon_1, and\ \epsilon_2$) as well as the PP reference energy, the $n$=1, 2, and 3 or the $n$=1, 2, 3, and 4 body iFCI energies for each molecule in this work.
- Geometry coordinates and 2D structures for each species in the isodesmic reaction set.
- Additional convergence results for 2-body MBBSA.

TOC:

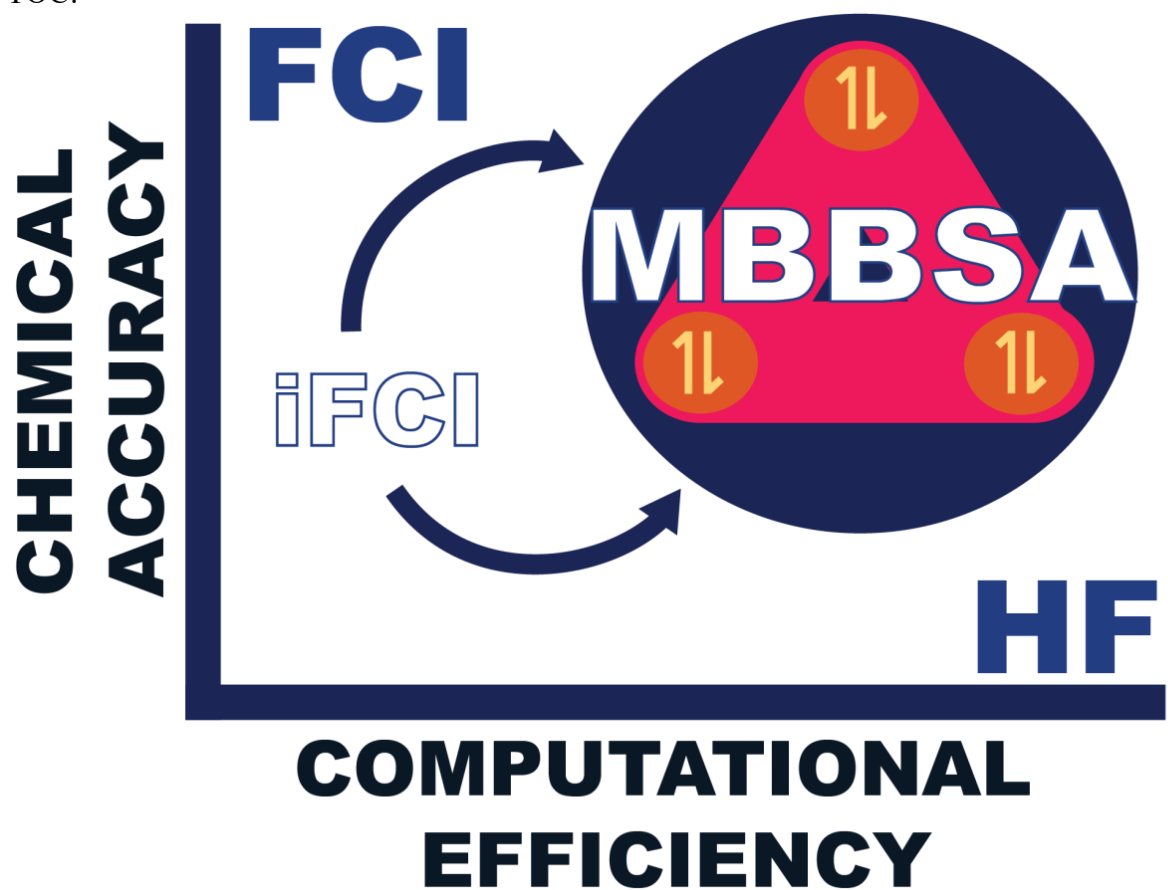

Supporting Information for
Many-Body Basis Set Amelioration Method
for Incremental Full Configuration Interaction

Jeffrey Hatch, Alan E Rask, Duy-Khoi Dang, Paul M. Zimmerman *
Department of Chemistry, University of Michigan
930 N. University Ave, Ann Arbor, MI 48109, USA, *paulzim@umich.edu

**Table of Contents:**

I.  **Small Molecule Energy Convergence:**

The perfect pairing orbitals in the QZ and 5Z basis sets for water were qualitatively different than the DZ and TZ basis sets, when the localized orbitals were formed using the default Pipek-Mezey localization procedure. The Boys localization procedure was more effective in this case and was therefore employed for water (all basis sets).[1]

**Table S1**. The $\zeta$, $\varepsilon_1$, and $\varepsilon_2$, basis set, and iFCI settings for each compound in the small molecule energy convergence studies. The $n = 1, 2, 3$ and $4$ energies from the corresponding iFCI calculation are also reported. Units of $\epsilon$ are in μHa.

| Compound | Basis | -log($\zeta$) | $\varepsilon_1$ | $\varepsilon_2$ | $E_{ref}$ | n = 1 | n = 2 | n = 3 | n = 4 |
|---|---|---|---|---|---|---|---|---|---|
| $H_2O$ | cc-pVDZ | 8.5 | 1000 | 0.1 | -76.090 | -76.120 | -76.243 | -76.242 | -76.243 |
| $H_2O$ | cc-pVTZ | 8.5 | 1000 | 0.1 | -76.123 | -76.173 | -76.334 | -76.334 | -76.335 |
| $H_2O$ | cc-pVQZ | 8.5 | 1000 | 0.1 | -76.132 | -76.188 | -76.361 | -76.362 | -76.362 |
| $H_2O$ | cc-pV5Z | 8.5 | 1000 | 0.1 | -76.135 | -76.193 | -76.370 | -76.371 | -76.371 |
| $CH_2$ | cc-pVDZ | 8.5 | 1000 | 0.1 | -38.947 | -38.974 | -39.039 | -39.042 | a |
| $CH_2$ | cc-pVTZ | 8.5 | 1000 | 0.1 | -38.959 | -38.997 | -39.076 | -39.079 | a |
| $CH_2$ | cc-pVQZ | 8.5 | 1000 | 0.1 | -38.962 | -39.002 | -39.085 | -39.088 | a |
| $CH_2$ | cc-pV5Z | 8.5 | 1000 | 0.1 | -38.962 | -39.004 | -39.088 | -39.091 | a |
| $CH_4$ | cc-pVDZ | 8.5 | 1000 | 0.1 | -40.258 | -40.306 | -40.384 | -40.387 | -40.388 |
| $CH_4$ | cc-pVTZ | 8.5 | 1000 | 0.1 | -40.274 | -40.339 | -40.435 | -40.439 | -40.439 |
| $CH_4$ | cc-pVQZ | 8.5 | 1000 | 0.1 | -40.278 | -40.347 | -40.447 | -40.452 | -40.452 |
| $CH_4$ | cc-pV5Z | 8.5 | 1000 | 0.1 | -40.279 | -40.350 | -40.451 | -40.455 | -40.455 |

a. $CH_2$ has 6 valence electrons, so $n = 4$ is not applicable.

II. Isodesmic Reaction Set:

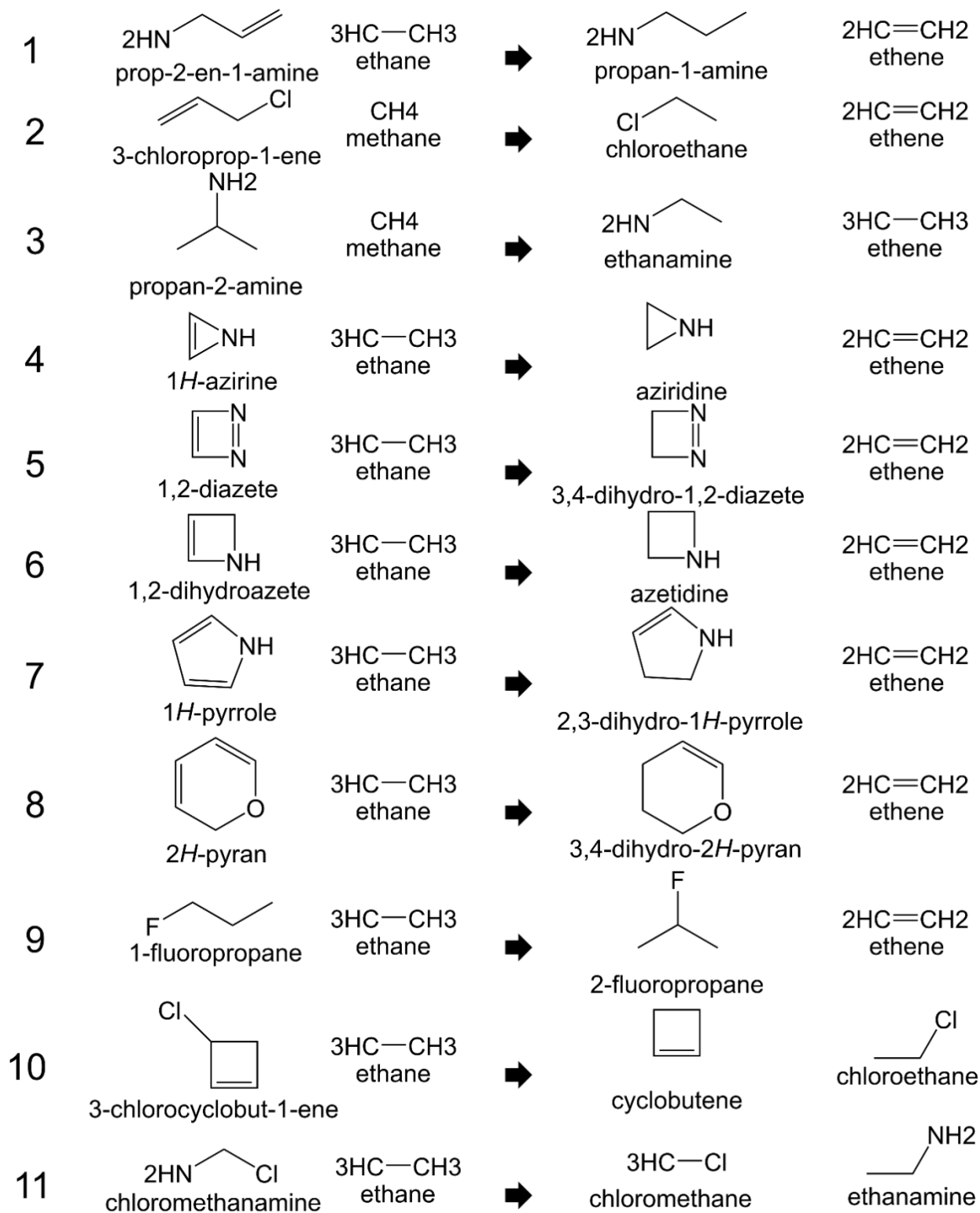

**Figure S1.** Lewis structures of the 11 isodesmic reactions and their IUPAC names.

**Table S2.** The ζ, ε₁, and ε₂, basis sets, and iFCI settings for each compound in the isodesmic reaction set. The n=1, 2, 3 and 4 energies from the corresponding iFCI calculation are also reported. Geometries of all chemical species can be found in Table S3. Units of $\epsilon$ are in μHa.

| Compound | Basis | -log(ζ) | ε₁ | ε₂ | E_ref (Ha) | n = 1 (Ha) | n = 2 (Ha) | n = 3 (Ha) |
|---|---|---|---|---|---|---|---|---|
| 1,2-diazete | cc-pVDZ | 8.5 | 500 | 0.1 | -185.808 | -185.875 | -186.231 | -186.268 |
| 1,2-diazete | cc-pVTZ | 8.5 | 500 | 0.1 | -185.862 | -185.967 | -186.403 | -186.445 |
| 1,2-dihydroazete | cc-pVDZ | 8.5 | 500 | 0.1 | -171.043 | -171.137 | -171.472 | -171.515 |
| 1,2-dihydroazete | cc-pVTZ | 8.5 | 500 | 0.1 | -171.097 | -171.235 | -171.646 | -171.695 |
| 1-fluoropropane | cc-pVDZ | 8.5 | 500 | 0.1 | -217.317 | -217.435 | -217.785 | -217.813 |
| 1-fluoropropane | cc-pVTZ | 8.5 | 500 | 0.1 | -217.395 | -217.573 | -218.017 | -218.048 |
| 1H-azirine | cc-pVDZ | 8.5 | 500 | 0.1 | -131.922 | -131.981 | -132.242 | -132.267 |
| 1H-azirine | cc-pVTZ | 8.5 | 500 | 0.1 | -131.964 | -132.056 | -132.377 | -132.405 |
| 1H-pyrrole | cc-pVDZ | 8.5 | 500 | 0.1 | -209.002 | -209.106 | -209.516 | -209.595 |
| 1H-pyrrole | cc-pVTZ | 8.5 | 500 | 0.1 | -209.064 | -209.219 | -209.725 | -209.808 |
| 2,3-dihydro-1H-pyrrole | cc-pVDZ | 8.5 | 500 | 0.1 | -210.176 | -210.300 | -210.707 | -210.763 |
| 2,3-dihydro-1H-pyrrole | cc-pVTZ | 8.5 | 500 | 0.1 | -210.241 | -210.423 | -210.925 | -210.985 |
| 2-fluoropropane | cc-pVDZ | 8.5 | 500 | 0.1 | -217.323 | -217.441 | -217.789 | -217.819 |
| 2-fluoropropane | cc-pVTZ | 8.5 | 500 | 0.1 | -217.401 | -217.579 | -218.021 | -218.055 |
| 2H-pyran | cc-pVDZ | 8.5 | 500 | 0.1 | -267.897 | -268.026 | -268.518 | -268.590 |
| 2H-pyran | cc-pVTZ | 8.5 | 500 | 0.1 | -267.980 | -268.175 | -268.786 | -268.864 |
| 3,4-dihydro-1,2-diazete | cc-pVDZ | 8.5 | 500 | 0.1 | -187.066 | -187.152 | -187.504 | -187.542 |
| 3,4-dihydro-1,2-diazete | cc-pVTZ | 8.5 | 500 | 0.1 | -187.122 | -187.255 | -187.685 | -187.728 |
| 3,4-dihydro-2H-pyran | cc-pVDZ | 8.5 | 500 | 0.1 | -269.104 | -269.252 | -269.745 | -269.812 |
| 3,4-dihydro-2H-pyran | cc-pVTZ | 8.5 | 500 | 0.1 | -269.190 | -269.410 | -270.019 | -270.093 |
| 3-chlorocyclobut-1-ene | cc-pVDZ | 8.5 | 500 | 0.1 | -614.023 | -614.139 | -614.539 | -614.591 |
| 3-chlorocyclobut-1-ene | cc-pVTZ | 8.5 | 500 | 0.1 | -614.088 | -614.264 | -614.763 | -614.828 |
| 3-chloro-prop-1-ene | cc-pVDZ | 8.5 | 500 | 0.1 | -576.155 | -576.259 | -576.580 | -576.616 |
| 3-chloro-prop-1-ene | cc-pVTZ | 8.5 | 500 | 0.1 | -576.213 | -576.368 | -576.773 | -576.820 |
| azetidine | cc-pVDZ | 8.5 | 500 | 0.1 | -172.275 | -172.387 | -172.721 | -172.754 |
| azetidine | cc-pVTZ | 8.5 | 500 | 0.1 | -172.330 | -172.494 | -172.902 | -172.940 |
| azridine | cc-pVDZ | 8.5 | 500 | 0.1 | -133.187 | -133.268 | -133.528 | -133.551 |
| azridine | cc-pVTZ | 8.5 | 500 | 0.1 | -133.231 | -133.350 | -133.669 | -133.696 |
| chloroethane | cc-pVDZ | 8.5 | 500 | 0.1 | -538.280 | -538.374 | -538.620 | -538.645 |
| chloroethane | cc-pVTZ | 8.5 | 500 | 0.1 | -538.327 | -538.466 | -538.780 | -538.816 |
| chloromethanamine | cc-pVDZ | 8.5 | 500 | 0.1 | -554.271 | -554.355 | -554.624 | -554.650 |
| chloromethanamine | cc-pVTZ | 8.5 | 500 | 0.1 | -554.326 | -554.455 | -554.803 | -554.839 |
| chloromethane | cc-pVDZ | 8.5 | 500 | 0.1 | -499.196 | -499.259 | -499.427 | -499.442 |
| chloromethane | cc-pVTZ | 8.5 | 500 | 0.1 | -499.231 | -499.327 | -499.547 | -499.571 |
| cyclobutene | cc-pVDZ | 8.5 | 500 | 0.1 | -155.079 | -155.180 | -155.491 | -155.525 |
| cyclobutene | cc-pVTZ | 8.5 | 500 | 0.1 | -155.125 | -155.271 | -155.647 | -155.686 |
| ethanamine | cc-pVDZ | 8.5 | 500 | 0.1 | -134.413 | -134.512 | -134.770 | -134.790 |
| ethanamine | cc-pVTZ | 8.5 | 500 | 0.1 | -134.460 | -134.602 | -134.920 | -134.944 |
| ethane | cc-pVDZ | 8.5 | 500 | 0.1 | -79.338 | -79.416 | -79.572 | -79.583 |
| ethane | cc-pVTZ | 8.5 | 500 | 0.1 | -79.366 | -79.474 | -79.664 | -79.677 |
| ethylene | cc-pVDZ | 8.5 | 500 | 0.1 | -78.135 | -78.193 | -78.346 | -78.357 |
| ethylene | cc-pVTZ | 8.5 | 500 | 0.1 | -78.161 | -78.244 | -78.430 | -78.443 |
| methane | cc-pVDZ | 8.5 | 500 | 0.1 | -40.258 | -40.305 | -40.384 | -40.387 |
| methane | cc-pVTZ | 8.5 | 500 | 0.1 | -40.274 | -40.339 | -40.435 | -40.439 |
| prop-2-en-1-amine | cc-pVDZ | 8.5 | 500 | 0.1 | -172.293 | -172.403 | -172.734 | -172.766 |
| prop-2-en-1-amine | cc-pVTZ | 8.5 | 500 | 0.1 | -172.350 | -172.510 | -172.918 | -172.955 |
| propan-1-amine | cc-pVDZ | 8.5 | 500 | 0.1 | -173.496 | -173.624 | -173.960 | -173.989 |
| propan-1-amine | cc-pVTZ | 8.5 | 500 | 0.1 | -173.552 | -173.735 | -174.146 | -174.181 |
| propan-2-amine | cc-pVDZ | 8.5 | 500 | 0.1 | -173.496 | -173.625 | -173.962 | -173.994 |
| propan-2-amine | cc-pVTZ | 8.5 | 500 | 0.1 | -173.555 | -173.740 | -174.152 | -174.189 |
| propane | cc-pVDZ | 8.5 | 500 | 0.1 | -118.419 | -118.527 | -118.762 | -118.782 |
| propane | cc-pVTZ | 8.5 | 500 | 0.1 | -118.459 | -118.610 | -118.894 | -118.918 |

### III. HTBH Reaction Set:

**Table S3:** The unsigned error of the thermodynamics ($\Delta E$) and the activation energy ($E_a$) for 18 reactions from the HTBH database, using the MBBSA-3 and MBBSA-4 methods.

| Reaction | Reactants | | Products | | $\Delta E$ MBBSA-3 (kcal/mol) | $\Delta E$ MBBSA-4 (kcal/mol) | $E_a$ MBBSA-3 (kcal/mol) | $E_a$ MBBSA-4 (kcal/mol) |
|---|---|---|---|---|---|---|---|---|
| 1 | H | HCl | $H_2$ | Cl | 0.16 | -0.09 | -1.04 | 0.19 |
| 2 | OH | $H_2$ | $H_2O$ | H | 1.37 | -0.06 | -1.24 | 0.41 |
| 3 | $CH_3$ | $H_2$ | $CH_4$ | H | -0.56 | 0.16 | -0.88 | 0.17 |
| 4 | OH | $CH_4$ | $H_2O$ | $CH_3$ | 1.93 | -0.26 | -1.75 | -0.11 |
| 5[a] | H | $H_2$ | $H_2$ | H | 0.00 | 0.00 | 0.00 | 0.00 |
| 6 | OH | $NH_3$ | $H_2O$ | $NH_2$ | 1.09 | -0.16 | -1.50 | -0.55 |
| 7 | HCl | $CH_3$ | $CH_4$ | Cl | -0.40 | 0.41 | -1.43 | 0.25 |
| 8 | OH | $C_2H_6$ | $H_2O$ | $C_2H_5$ | 0.62 | 0.08 | -3.76 | 0.66 |
| 9 | F | $H_2$ | HF | H | 0.09 | 0.01 | -2.24 | 0.47 |
| 10 | O | $CH_4$ | OH | $CH_3$ | -0.45 | -0.04 | 0.37 | 0.25 |
| 11 | H | $PH_3$ | $H_2$ | $PH_2$ | 0.18 | -0.09 | -0.41 | 0.07 |
| 12 | H | OH | $H_2$ | O | 1.01 | -0.16 | 1.84 | -0.19 |
| 13 | H | $H_2S$ | $H_2$ | HS | 0.17 | -0.22 | -0.56 | 0.18 |
| 14 | O | HCl | OH | Cl | -0.85 | 0.06 | -0.85 | -0.65 |
| 15 | $CH_3$ | $NH_2$ | $CH_4$ | NH | 0.72 | 0.14 | -1.10 | -0.28 |
| 16 | $C_2H_5$ | $NH_2$ | $C_2H_6$ | NH | 2.03 | 0.15 | -0.83 | 0.03 |

| | 17 | NH$_2$ | C$_2$H$_6$ | NH$_3$ | C$_2$H$_5$ | -0.48 | 0.26 | -1.97 | 0.81 |
|---|---|---|---|---|---|---|---|---|---|
| | 18 | NH$_2$ | CH$_4$ | NH$_3$ | CH$_3$ | 0.84 | -0.07 | -0.95 | -0.18 |
| | MUE | | | | | 0.72 | 0.14 | 1.26 | 0.30 |

[a] $E^{iFCI}[n=2]$ is exact for the 1, 2 and 3-electrons systems in reaction 5.

**Table S4**. The $\zeta$, $\varepsilon_1$, $\varepsilon_2$, and spin iFCI settings for each compound in the DZ and TZ bases used in the HTBH reaction set. The n=1, 2, 3 and 4 energies from the corresponding iFCI calculation are also reported. Units of $\epsilon$ are in µHa.

| Compound | Basis[a] | Spin[b] | -log($\zeta$) | $\varepsilon_1$ | $\varepsilon_2$ | E$_{ref}$ (Ha) | n = 1 (Ha) | n = 2 (Ha) | n = 3 (Ha) | n = 4 (Ha) |
|---|---|---|---|---|---|---|---|---|---|---|
| C$_2$H$_5$ | DZ | 2 | 6.5 | 500 | 0.01 | -78.695 | -78.764 | -78.910 | -78.929 | -78.928 |
| C$_2$H$_5$ | TZ | 2 | 6.5 | 500 | 0.01 | -78.722 | -78.816 | -78.994 | -79.015 | -79.013 |
| C$_2$H$_6$ | DZ | 1 | 6.5 | 500 | 0.01 | -79.341 | -79.423 | -79.587 | -79.598 | -79.597 |
| C$_2$H$_6$ | TZ | 1 | 6.5 | 500 | 0.01 | -79.367 | -79.474 | -79.663 | -79.676 | -79.674 |
| CH$_4$ | DZ | 1 | 6.5 | 500 | 0.01 | -40.258 | -40.306 | -40.384 | -40.387 | -40.388 |
| CH$_4$ | TZ | 1 | 6.5 | 500 | 0.01 | -40.274 | -40.339 | -40.435 | -40.439 | -40.439 |
| CH$_3$ | DZ | 2 | 6.5 | 500 | 0.01 | -39.614 | -39.652 | -39.721 | -39.729 | -39.728 |
| CH$_3$ | TZ | 2 | 6.5 | 500 | 0.01 | -39.630 | -39.682 | -39.766 | -39.775 | -39.774 |
| H$_2$O | DZ | 1 | 6.5 | 500 | 0.01 | -76.086 | -76.117 | -76.243 | -76.242 | -76.242 |
| H$_2$O | TZ | 1 | 6.5 | 500 | 0.01 | -76.119 | -76.168 | -76.334 | -76.333 | -76.334 |
| H$_2$S | DZ | 1 | 6.5 | 500 | 0.01 | -398.736 | -398.772 | -398.857 | -398.866 | -398.866 |
| H$_2$S | TZ | 1 | 6.5 | 500 | 0.01 | -398.757 | -398.812 | -398.924 | -398.938 | -398.937 |
| HS | DZ | 2 | 6.5 | 500 | 0.01 | -398.115 | -398.141 | -398.221 | -398.230 | -398.229 |
| HS | TZ | 2 | 6.5 | 500 | 0.01 | -398.131 | -398.173 | -398.280 | -398.292 | -398.291 |
| HCl | DZ | 1 | 6.5 | 500 | 0.01 | -460.123 | -460.156 | -460.249 | -460.256 | -460.255 |
| HCl | TZ | 1 | 6.5 | 500 | 0.01 | -460.144 | -460.197 | -460.326 | -460.340 | -460.339 |
| CL | DZ | 2 | 6.5 | 500 | 0.01 | -459.487 | -459.513 | -459.595 | -459.601 | -459.601 |
| CL | TZ | 2 | 6.5 | 500 | 0.01 | -459.502 | -459.543 | -459.662 | -459.674 | -459.674 |
| HF | DZ | 1 | 6.5 | 500 | 0.01 | -100.077 | -100.103 | -100.232 | -100.229 | -100.229 |
| HF | TZ | 1 | 6.5 | 500 | 0.01 | -100.119 | -100.167 | -100.342 | -100.339 | -100.339 |
| F | DZ | 2 | 6.5 | 500 | 0.01 | -99.451 | -99.482 | -99.551 | -99.570 | -99.568 |
| F | TZ | 2 | 6.5 | 500 | 0.01 | -99.484 | -99.533 | -99.641 | -99.661 | -99.658 |
| NH$_3$ | DZ | 1 | 6.5 | 500 | 0.01 | -56.258 | -56.296 | -56.401 | -56.403 | -56.403 |
| NH$_3$ | TZ | 1 | 6.5 | 500 | 0.01 | -56.283 | -56.338 | -56.472 | -56.474 | -56.474 |
| NH$_2$ | DZ | 2 | 6.5 | 500 | 0.01 | -55.643 | -55.678 | -55.747 | -55.766 | -55.764 |
| NH$_2$ | TZ | 2 | 6.5 | 500 | 0.01 | -55.664 | -55.715 | -55.807 | -55.827 | -55.825 |
| NH | DZ | 3 | 6.5 | 500 | 0.01 | -54.984 | -55.007 | -55.089 | -55.092 | -55.092 |
| NH | TZ | 3 | 6.5 | 500 | 0.01 | -54.999 | -55.033 | -55.139 | -55.141 | -55.141 |
| OH | DZ | 2 | 6.5 | 500 | 0.01 | -75.492 | -75.528 | -75.584 | -75.616 | -75.612 |
| OH | TZ | 2 | 6.5 | 500 | 0.01 | -75.519 | -75.569 | -75.660 | -75.694 | -75.690 |
| **O** | DZ | 3 | 6.5 | 500 | 0.01 | -74.800 | -74.821 | -74.902 | -74.910 | -74.910 |
| **O** | TZ | 3 | 6.5 | 500 | 0.01 | -74.820 | -74.848 | -74.958 | -74.967 | -74.967 |
| **PH$_2$** | DZ | 2 | 6.5 | 500 | 0.01 | -341.902 | -341.934 | -342.000 | -342.012 | -342.011 |
| **PH$_2$** | TZ | 2 | 6.5 | 500 | 0.01 | -341.917 | -341.962 | -342.048 | -342.061 | -342.059 |
| **PH$_3$** | DZ | 1 | 6.5 | 500 | 0.01 | -342.517 | -342.561 | -342.628 | -342.639 | -342.638 |
| **PH$_3$** | TZ | 1 | 6.5 | 500 | 0.01 | -342.536 | -342.598 | -342.683 | -342.694 | -342.693 |
| **H** | DZ | 2 | 6.5 | 500 | 0.01 | -0.499 | -0.499 | -0.499 | -0.499 | -0.499 |
| **H** | TZ | 2 | 6.5 | 500 | 0.01 | -0.500 | -0.500 | -0.500 | -0.500 | -0.500 |
| **H2** | DZ | 1 | 6.5 | 500 | 0.01 | -1.164 | -1.164 | -1.164 | -1.164 | -1.164 |
| **H2** | TZ | 1 | 6.5 | 500 | 0.01 | -1.172 | -1.172 | -1.172 | -1.172 | -1.172 |
| **TS1** | DZ | 2 | 6.5 | 500 | 0.01 | -460.587 | -460.623 | -460.731 | -460.746 | -460.745 |
| **TS1** | TZ | 2 | 6.5 | 500 | 0.01 | -460.608 | -460.661 | -460.809 | -460.832 | -460.830 |
| **TS2** | DZ | 2 | 6.5 | 500 | 0.01 | -76.573 | -76.614 | -76.727 | -76.741 | -76.745 |
| **TS2** | TZ | 2 | 6.5 | 500 | 0.01 | -76.603 | -76.664 | -76.812 | -76.831 | -76.834 |
| **TS3** | DZ | 2 | 6.5 | 500 | 0.01 | -40.709 | -40.755 | -40.855 | -40.869 | -40.868 |
| **TS3** | TZ | 2 | 6.5 | 500 | 0.01 | -40.726 | -40.788 | -40.907 | -40.923 | -40.922 |
| **TS4** | DZ | 2 | 6.5 | 500 | 0.01 | -115.672 | -115.743 | -115.929 | -115.969 | -115.962 |
| **TS4** | TZ | 2 | 6.5 | 500 | 0.01 | -115.714 | -115.816 | -116.055 | -116.102 | -116.095 |
| **TS5** | DZ | 2 | 6.5 | 500 | 0.01 | -1.600 | -1.611 | -1.686 | -1.686 | -1.686 |
| **TS5** | TZ | 2 | 6.5 | 500 | 0.01 | -1.604 | -1.618 | -1.715 | -1.715 | -1.715 |
| **TS6** | DZ | 2 | 6.5 | 500 | 0.01 | -131.673 | -131.735 | -131.941 | -132.003 | -131.983 |
| **TS6** | TZ | 2 | 6.5 | 500 | 0.01 | -131.722 | -131.818 | -132.086 | -132.153 | -132.134 |
| **TS7** | DZ | 2 | 6.5 | 500 | 0.01 | -499.705 | -499.770 | -499.946 | -499.980 | -499.976 |
| **TS7** | TZ | 2 | 6.5 | 500 | 0.01 | -499.739 | -499.837 | -500.067 | -500.110 | -500.105 |
| **TS8** | DZ | 2 | 6.5 | 500 | 0.01 | -154.762 | -154.865 | -155.122 | -155.171 | -155.164 |
| **TS8** | TZ | 2 | 6.5 | 500 | 0.01 | -154.816 | -154.963 | -155.288 | -155.346 | -155.338 |
| **TS9** | DZ | 2 | 6.5 | 500 | 0.01 | -100.568 | -100.610 | -100.707 | -100.715 | -100.721 |
| **TS9** | TZ | 2 | 6.5 | 500 | 0.01 | -100.604 | -100.668 | -100.807 | -100.819 | -100.823 |
| **TS10** | DZ | 3 | 6.5 | 500 | 0.01 | -114.995 | -115.113 | -115.253 | -115.270 | -115.268 |
| **TS10** | TZ | 3 | 6.5 | 500 | 0.01 | -115.035 | -115.201 | -115.373 | -115.392 | -115.390 |
| **TS11** | DZ | 2 | 6.5 | 500 | 0.01 | -342.995 | -343.036 | -343.114 | -343.132 | -343.130 |
| **TS11** | TZ | 2 | 6.5 | 500 | 0.01 | -343.013 | -343.071 | -343.170 | -343.190 | -343.187 |
| **TS12** | DZ | 3 | 6.5 | 500 | 0.01 | -75.874 | -75.902 | -76.021 | -76.045 | -76.044 |
| **TS12** | TZ | 3 | 6.5 | 500 | 0.01 | -75.900 | -75.937 | -76.097 | -76.120 | -76.119 |
| **TS13** | DZ | 2 | 6.5 | 500 | 0.01 | -399.208 | -399.242 | -399.340 | -399.359 | -399.357 |
| **TS13** | TZ | 2 | 6.5 | 500 | 0.01 | -399.229 | -399.281 | -399.409 | -399.433 | -399.430 |
| **TS14** | DZ | 3 | 6.5 | 500 | 0.01 | -534.866 | -534.914 | -535.105 | -535.137 | -535.142 |
| **TS14** | TZ | 3 | 6.5 | 500 | 0.01 | -534.908 | -534.987 | -535.247 | -535.287 | -535.293 |
| **TS15** | DZ | 3 | 6.5 | 500 | 0.01 | -95.176 | -95.240 | -95.420 | -95.452 | -95.455 |

| | | | | | | | | | |
|---|---|---|---|---|---|---|---|---|---|
| **TS15** | TZ | 3 | 6.5 | 500 | 0.01 | -95.210 | -95.300 | -95.522 | -95.557 | -95.560 |
| **TS16** | DZ | 3 | 6.5 | 500 | 0.01 | -134.253 | -134.346 | -134.607 | -134.647 | -134.650 |
| **TS16** | TZ | 3 | 6.5 | 500 | 0.01 | -134.298 | -134.430 | -134.749 | -134.794 | -134.796 |
| **TS17** | DZ | 2 | 6.5 | 500 | 0.01 | -134.904 | -135.006 | -135.269 | -135.322 | -135.315 |
| **TS17** | TZ | 2 | 6.5 | 500 | 0.01 | -134.950 | -135.095 | -135.418 | -135.476 | -135.468 |
| **TS18** | DZ | 2 | 6.5 | 500 | 0.01 | -95.818 | -95.889 | -96.078 | -96.119 | -96.114 |
| **TS18** | TZ | 2 | 6.5 | 500 | 0.01 | -95.853 | -95.953 | -96.188 | -96.233 | -96.228 |

[a] DZ represents the cc-pVDZ basis and TZ represents the cc-pVTZ basis.
[b] 1 refers to a singlet, 2 to a doublet and 3 to a triplet spin state.

The iFCI/MBBSA code formally only treats systems with even numbers of electrons. This is not a hindrance, however, as any odd-electron count can be converted into an even electron count by adding a hydrogen atom (with one electron) to an arbitrary location. Therefore for each doublet state molecule, a lone hydrogen atom was placed 6 Å from the molecule, and a triplet spin state was assigned, guaranteeing the extra electron is localized on the hydrogen atom. This distance was sufficiently far such that the lone electron did not interact with the remainder of the system. After subtraction of the energy of the extra hydrogen atom, this method gives the energy of the doublet state of the primary molecule.

### IV. Automerization of Cyclobutadiene and Criegee Intermediate Reaction:

**Table S5.** The $\zeta$, $\varepsilon_1$, $\varepsilon_2$, basis set, and iFCI settings for each geometry involved in the automerization of cyclobutadiene. The n=1, 2, and 3 energies from the corresponding iFCI calculation are also reported. Units of $\epsilon$ are in µHa.

| Geometry | Basis | -log($\zeta$) | $\varepsilon_1$ | $\varepsilon_2$ | $E_{ref}$ | n = 1 | n = 2 | n = 3 |
|---|---|---|---|---|---|---|---|---|
| $D_{2h}$ | cc-pVDZ | 10.5 | 500 | 0.1 | -153.831 | -153.914 | -154.242 | -154.277 |
| $D_{2h}$ | cc-pVTZ | 10.5 | 500 | 0.1 | -153.869 | -153.989 | -154.374 | -154.412 |
| $D_{4h}$ | cc-pVDZ | 10.5 | 500 | 0.1 | -153.793 | -153.876 | -154.231 | -154.265 |
| $D_{4h}$ | cc-pVTZ | 10.5 | 500 | 0.1 | -153.830 | -153.949 | -154.361 | -154.399 |

The significant differences in the correlation energy for butadiene at the iFCI(n=2) level result in large MBBSA. At minimum the iFCI(n=3) level is required to achieve the convergence necessary for MBBSA to achieve the threshold for chemical accuracy.

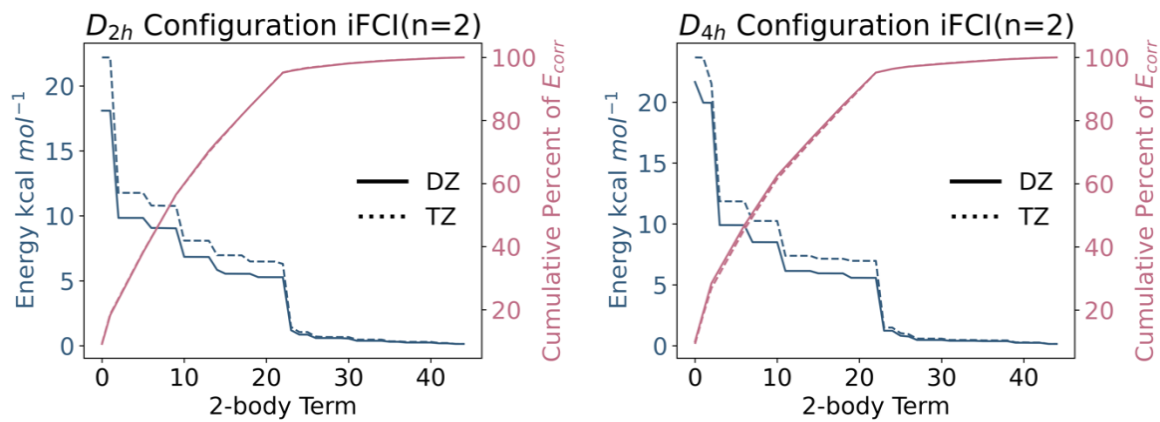

**Figure S2.** A comparison of the magnitude of correlation energy of butadiene in each basis and the cumulative percent of the total correlation energy recovered by each 2-body term.

**Table S6.** The $\zeta$, $\varepsilon_1$, $\varepsilon_2$, basis set, and iFCI settings for each geometry in the Criegee intermediate reaction. The n=1, 2, and 3 energies from the corresponding MBBSA iFCI calculation are also reported. Units of $\epsilon$ are in µHa.

| Compound | Basis | -log($\zeta$) | $\varepsilon_1$ | $\varepsilon_2$ | $E_{ref}$ | n = 1 | n = 2 | n = 3 |
|---|---|---|---|---|---|---|---|---|
| ethylene | cc-pVDZ | 10.5 | 500 | 0.1 | -78.136 | -78.194 | -78.347 | -78.359 |
| ethylene | cc-pVTZ | 10.5 | 500 | 0.1 | -78.162 | -78.244 | -78.430 | -78.444 |
| dioxirane | cc-pVDZ | 10.5 | 500 | 0.1 | -188.780 | -188.840 | -189.145 | -189.170 |
| dioxirane | cc-pVTZ | 10.5 | 500 | 0.1 | -188.848 | -188.947 | -189.336 | -189.368 |
| Product | cc-pVDZ | 10.5 | 500 | 0.1 | -267.040 | -267.162 | -267.608 | -267.660 |
| Product | cc-pVTZ | 10.5 | 500 | 0.1 | -267.132 | -267.321 | -267.883 | -267.943 |
| TS | cc-pVDZ | 10.5 | 500 | 0.1 | -266.849 | -266.964 | -267.423 | -267.500 |
| TS | cc-pVTZ | 10.5 | 500 | 0.1 | -266.938 | -267.115 | -267.695 | -267.784 |

**Table S7.** The xyz coordinates (Å) for each molecule in the isodesmic reaction set.

| 1,2-diazete | | | |
|---|---|---|---|
| C | -1.29974 | 2.415476 | -0.01641 |
| C | -1.29637 | 3.745031 | -0.03286 |
| N | -2.81842 | 3.717259 | -0.24291 |
| N | -2.82165 | 2.445759 | -0.22718 |
| H | -0.62215 | 1.589326 | 0.087769 |
| H | -0.61459 | 4.570058 | 0.050895 |
| 1,2-dihydroazete | | | |
| C | -1.36947 | 2.427449 | 0.007027 |
| C | -1.30752 | 3.780357 | -0.00874 |
| C | -2.82526 | 3.810234 | -0.03237 |
| **H** | -0.53755 | 4.520208 | -0.0061 |
| **N** | -2.74889 | 2.349421 | -0.01312 |
| **H** | -3.26936 | 4.209464 | -0.93865 |
| **H** | -3.29675 | 4.231597 | 0.849732 |
| **H** | -3.42084 | 1.613614 | -0.01427 |
| **H** | -0.67268 | 1.608036 | 0.027829 |
| **1-fluoropropane** | | | |
| **C** | -2.42367 | 1.52852 | 0.004162 |
| **C** | -1.16388 | 2.372726 | 0.048139 |
| **H** | -2.48498 | 0.966852 | -0.92345 |
| **H** | -3.31099 | 2.146769 | 0.076837 |
| **H** | -2.44589 | 0.817442 | 0.824878 |
| **C** | 0.063387 | 1.511912 | -0.05286 |
| **H** | -1.1181 | 2.938777 | 0.973424 |
| **H** | -1.15705 | 3.087668 | -0.76885 |
| **F** | 1.206758 | 2.292081 | -0.01176 |
| **H** | 0.071883 | 0.957179 | -0.98715 |
| **H** | 0.111227 | 0.806804 | 0.772499 |
| **1H-azirine** | | | |
| **C** | -1.25637 | 2.573852 | 0.031871 |
| **N** | -2.73285 | 2.656068 | -0.30147 |
| **C** | -1.71717 | 3.75916 | -0.08197 |
| **H** | -3.23204 | 2.547181 | 0.581659 |
| **H** | -0.50614 | 1.821269 | 0.11784 |
| **H** | -1.67211 | 4.820179 | -0.17449 |
| **1H-pyrrole** | | | |
| **C** | -1.59949 | 2.383985 | -0.01626 |
| **C** | -0.19183 | 2.477629 | -0.01485 |
| **C** | 0.306996 | 1.197471 | -0.10715 |
| **C** | -1.92416 | 1.049044 | -0.10938 |
| **N** | -0.75685 | 0.346439 | -0.16335 |
| **H** | -0.69046 | -0.6505 | -0.23404 |
| **H** | -2.30045 | 3.191598 | 0.043403 |
| **H** | -2.87365 | 0.55532 | -0.14075 |
| **H** | 0.395856 | 3.370967 | 0.046103 |
| **H** | 1.313578 | 0.833873 | -0.13656 |
| **2,3-dihydro-1H-pyrrole** | | | |
| **C** | -1.62332 | 2.469013 | -0.01515 |
| **C** | -0.11639 | 2.470901 | -0.01835 |
| **C** | 0.320186 | 1.20952 | -0.10511 |
| **C** | -1.98932 | 0.965043 | -0.11658 |
| **N** | -0.70075 | 0.296641 | -0.16495 |
| **H** | -0.57945 | -0.69054 | -0.23231 |
| **H** | 0.500716 | 3.343525 | 0.039627 |
| **H** | 1.339418 | 0.869895 | -0.13046 |
| **H** | -2.03454 | 3.026519 | -0.85225 |
| **H** | -2.0307 | 2.907985 | 0.891487 |
| **H** | -2.57331 | 0.761134 | -1.01034 |
| **H** | -2.56933 | 0.64186 | 0.743991 |
| **2-fluoropropane** | | | |
| **C** | -2.43723 | 1.654474 | 0.079908 |
| **C** | -1.17352 | 2.462796 | 0.177723 |
| **H** | -2.48105 | 1.157949 | -0.88396 |
| **H** | -3.30609 | 2.293865 | 0.183142 |
| **H** | -2.46374 | 0.901661 | 0.861227 |
| **C** | 0.07331 | 1.647106 | -0.02225 |
| **F** | -1.21089 | 3.44588 | -0.81171 |
| **H** | -1.13331 | 2.987033 | 1.12816 |
| **H** | 0.035764 | 1.150562 | -0.98637 |
| **H** | 0.158798 | 0.893964 | 0.754515 |
| **H** | 0.95142 | 2.28137 | 0.0099 |
| **2H-pyran** | | | |
| **C** | -2.59854 | 2.408512 | -0.00045 |
| **C** | -1.16427 | 2.598738 | -0.01096 |
| **C** | -0.37887 | 1.517278 | 0.003009 |
| **C** | -3.10003 | 1.172467 | 0.022693 |
| **C** | -2.22394 | -0.03446 | 0.038284 |
| **O** | -0.81132 | 0.240927 | 0.026467 |
| **H** | -2.42667 | -0.66597 | -0.8264 |
| **H** | -2.41769 | -0.63682 | 0.925527 |
| **H** | -3.25498 | 3.260565 | -0.01126 |
| **H** | -4.16192 | 0.994757 | 0.031094 |
| **H** | -0.7161 | 3.572272 | -0.02946 |
| **H** | 0.69627 | 1.558772 | -0.00319 |
| **3,4-dihydro-1,2-diazete** | | | |
| **C** | -1.29738 | 2.316337 | 0.018957 |
| **C** | -1.28433 | 3.84417 | -0.06624 |
| **N** | -2.74494 | 3.718319 | -0.26093 |
| **N** | -2.75582 | 2.44463 | -0.18991 |
| **H** | -1.05249 | 1.871902 | 0.97606 |
| **H** | -0.80971 | 1.771867 | -0.78059 |
| **H** | -0.78821 | 4.287986 | -0.9209 |
| **H** | -1.03098 | 4.388021 | 0.835747 |
| **3,4-dihydro-2H-pyran** | | | |
| C | -2.63754 | 2.442055 | -0.19434 |
| C | -1.15266 | 2.520861 | -0.05376 |
| C | -0.40967 | 1.450428 | 0.234682 |
| C | -3.11257 | 1.104956 | 0.349452 |
| C | -2.19327 | 0.014453 | -0.14804 |
| O | -0.86733 | 0.179173 | 0.347242 |
| H | -2.15811 | 0.029786 | -1.23831 |
| H | -2.50682 | -0.96826 | 0.180954 |
| H | -0.64859 | 3.466194 | -0.13964 |
| H | 0.650117 | 1.506324 | 0.412413 |
| H | -2.93623 | 2.552024 | -1.23615 |
| H | -3.1066 | 3.258039 | 0.349417 |
| H | -4.13369 | 0.890932 | 0.048539 |
| H | -3.07815 | 1.113396 | 1.436034 |
| **3-chlorocyclobut-1-ene** | | | |
| C | -1.61854 | 2.345591 | 0.164573 |
| C | -0.15677 | 2.364358 | -0.15145 |
| C | -0.21857 | 0.812558 | -0.13038 |
| C | -1.69489 | 1.008585 | 0.105132 |
| H | 0.325499 | 0.388327 | 0.705167 |
| H | 0.074396 | 0.319558 | -1.05215 |
| Cl | 0.901299 | 3.112083 | 1.069639 |
| H | 0.115787 | 2.792148 | -1.10722 |
| H | -2.51282 | 0.31374 | 0.178908 |
| H | -2.32539 | 3.1375 | 0.333748 |
| **3-chloroprop-1-ene** | | | |
| C | -1.22208 | 1.343408 | 0.012901 |
| C | 0.080612 | 1.601797 | -0.00422 |
| H | 0.801588 | 0.804251 | 0.018923 |
| H | 0.455505 | 2.611707 | -0.0416 |
| C | -2.26168 | 2.419553 | -0.01791 |
| H | -1.5748 | 0.327189 | 0.050358 |
| Cl | -3.90817 | 1.745939 | 0.015711 |
| H | -2.18642 | 3.014601 | -0.91949 |
| H | -2.17545 | 3.075555 | 0.839315 |
| **azetidine** | | | |
| C | -1.34088 | 2.329719 | 0.043575 |
| C | -1.29293 | 3.856155 | -0.0985 |
| C | -2.82335 | 3.782594 | -0.02835 |
| N | -2.72583 | 2.375877 | -0.46611 |
| H | -3.4014 | 4.427113 | -0.68023 |
| H | -3.17137 | 3.896775 | 0.99881 |
| H | -3.36957 | 1.741451 | -0.01318 |
| H | -1.27105 | 2.034394 | 1.091011 |
| H | -0.6541 | 1.734628 | -0.54693 |
| H | -0.95915 | 4.147858 | -1.08598 |
| H | -0.77202 | 4.425195 | 0.659135 |
| **aziridine** | | | |
| C | -1.12184 | 2.524235 | -0.03698 |
| N | -2.55974 | 2.687715 | -0.27314 |
| C | -1.71766 | 3.872075 | -0.07759 |
| H | -3.05799 | 2.493308 | 0.58522 |
| H | -0.56951 | 2.16488 | -0.88665 |
| H | -0.79688 | 2.152233 | 0.919465 |
| H | -1.79508 | 4.410325 | 0.851423 |
| H | -1.59241 | 4.478869 | -0.95638 |
| **chloroethane** | | | |
| C | -1.23806 | 1.326704 | 0.013579 |
| C | -2.25729 | 2.434104 | -0.01842 |
| Cl | -3.91552 | 1.776486 | 0.014482 |
| H | -2.17837 | 3.026189 | -0.92026 |
| H | -2.16757 | 3.086972 | 0.839392 |
| H | -1.35683 | 0.673541 | -0.84316 |
| H | -1.34603 | 0.734264 | 0.914737 |
| H | -0.23778 | 1.749961 | -0.00718 |
| **chloromethanamine** | | | |
| C | -1.7724 | 1.114867 | 0.066025 |
| Cl | -0.17404 | 0.360267 | -0.22949 |
| N | -2.81188 | 0.139519 | -0.02107 |
| H | -1.82555 | 1.928883 | -0.64828 |
| H | -1.73895 | 1.515484 | 1.068915 |
| H | -3.06495 | -0.04859 | -0.97785 |
| H | -3.63126 | 0.44804 | 0.477851 |
| **chloromethane** | | | |
| C | -1.44909 | 2.008565 | 3.31E-08 |
| Cl | 0.28371 | 2.385309 | 7.85E-07 |
| H | -1.57895 | 0.980884 | 0.30559 |
| H | -1.83396 | 2.153779 | -0.99858 |
| H | -1.94641 | 2.670993 | 0.692992 |
| **cyclobutene** | | | |
| C | -3.37617 | 2.175131 | 0.005518 |
| C | -1.87646 | 2.319462 | 0.034664 |
| C | -1.8404 | 0.760565 | 0.036682 |
| H | -1.46566 | 2.781088 | 0.927052 |
| H | -1.43107 | 2.779651 | -0.84172 |
| C | -3.34519 | 0.83533 | 0.00766 |
| H | -1.40843 | 0.320689 | 0.930049 |
| H | -1.37447 | 0.319238 | -0.83873 |
| H | -4.17486 | 2.896358 | -0.01082 |
| H | -4.10969 | 0.077909 | -0.00627 |
| **ethanamine** | | | |
| C | -1.32057 | 1.329844 | 0.004407 |
| C | -2.38376 | 2.402954 | -0.04056 |
| N | -3.70347 | 1.790004 | 0.048775 |
| H | -2.24268 | 3.013678 | -0.93469 |
| H | -2.26699 | 3.061044 | 0.815163 |
| H | -1.41348 | 0.66512 | -0.8505 |
| H | -1.42915 | 0.736194 | 0.905266 |
| H | -0.32554 | 1.762141 | -0.01642 |
| H | -3.85023 | 1.192767 | -0.75311 |
| H | -4.42259 | 2.497594 | 0.008048 |
| **ethane** | | | |
| C | -0.7938 | 1.305533 | -1.9E-06 |
| C | 0.724009 | 1.305532 | -5.3E-07 |
| H | -1.18528 | 0.651255 | -0.77186 |
| H | -1.18528 | 2.301124 | -0.18069 |
| H | -1.18528 | 0.964218 | 0.95255 |
| H | 1.11549 | 1.646846 | -0.95255 |
| H | 1.115489 | 0.309941 | 0.180691 |
| H | 1.115489 | 1.959811 | 0.771861 |
| **ethylene** | | | |
| C | -0.69886 | 1.30553 | -4.4E-06 |
| C | 0.629061 | 1.30553 | -4.4E-06 |
| H | -1.25791 | 0.392885 | 0.114431 |
| H | -1.25791 | 2.218175 | -0.11444 |
| H | 1.188115 | 0.392885 | 0.114431 |
| H | 1.188115 | 2.218175 | -0.11444 |
| **methane** | | | |
| C | -1.01207 | 1.61886 | -2.2E-07 |
| H | 0.070363 | 1.618859 | 7.2E-08 |
| H | -1.37288 | 2.202676 | -0.83704 |
| H | -1.37288 | 2.05185 | 0.924118 |
| H | -1.37288 | 0.602055 | -0.08708 |
| **prop-2-en-1-amine** | | | |
| C | -1.12466 | 1.740023 | 0.339613 |
| C | -2.37497 | 2.538574 | 0.22335 |
| N | -3.44916 | 1.688053 | -0.27935 |
| H | -2.17518 | 3.424278 | -0.38367 |
| H | -2.67163 | 2.883889 | 1.211477 |
| H | -3.23751 | 1.408548 | -1.22664 |
| H | -4.31571 | 2.204984 | -0.31552 |
| C | 0.014337 | 2.050976 | -0.27101 |
| H | -1.18822 | 0.85055 | 0.945815 |
| H | 0.086233 | 2.930332 | -0.89048 |
| H | 0.895666 | 1.443064 | -0.16286 |
| **propan-1-amine** | | | |
| C | -1.68965 | 3.09127 | -0.04782 |
| C | -0.42772 | 2.22522 | -0.09318 |
| C | 0.83155 | 3.07679 | 0.03786 |
| H | -0.38929 | 1.67299 | -1.05732 |
| H | -0.45509 | 1.48835 | 0.73892 |
| H | 0.89089 | 3.80853 | -0.79594 |
| H | 1.72836 | 2.42305 | 0.00149 |
| H | 0.82923 | 3.62411 | 1.00457 |
| N | -2.88045 | 2.25644 | -0.16531 |
| H | -1.67134 | 3.83405 | -0.87674 |
| H | -1.72621 | 3.6387 | 0.9197 |
| H | -2.88358 | 1.79193 | -1.10274 |
| H | -3.7239 | 2.87234 | -0.11268 |
| **propan-2-amine** | | | |
| C | -1.27964 | 1.172275 | 0.358362 |
| C | 0.017094 | 0.477645 | -0.02023 |
| H | 0.06379 | -0.5192 | 0.408864 |
| H | 0.884612 | 1.035762 | 0.322194 |
| H | 0.081146 | 0.382687 | -1.10224 |
| C | -1.34731 | 2.558313 | -0.2449 |
| H | -0.50447 | 3.1649 | 0.070343 |
| H | -2.2685 | 3.052989 | 0.040697 |
| H | -1.32055 | 2.48855 | -1.33092 |
| N | -2.47767 | 0.440323 | -0.04213 |
| H | -1.32127 | 1.265846 | 1.441153 |
| H | -2.43823 | -0.50425 | 0.316995 |
| H | -2.4833 | 0.352796 | -1.05102 |
| propane | | | |
| C | -1.617 | -1.29529 | 0.011846 |
| C | -0.31562 | -0.51831 | 0.071454 |
| H | -2.47975 | -0.64348 | 0.103567 |
| H | -1.66333 | -2.02775 | 0.813061 |
| H | -1.70053 | -1.83185 | -0.92927 |
| C | 0.894023 | -1.42481 | -0.05461 |
| H | -0.2625 | 0.030973 | 1.007915 |
| H | -0.29835 | 0.223768 | -0.7226 |
| H | 0.908243 | -2.15891 | 0.746312 |
| H | 1.822804 | -0.86514 | -0.01259 |
| H | 0.870987 | -1.96622 | -0.99638 |